\documentclass[fleqn,twoside]{article}
\usepackage{espcrc2}

\usepackage{graphicx}

\newcommand{\AmS}{{\protect\the\textfont2
  A\kern-.1667em\lower.5ex\hbox{M}\kern-.125emS}}
\newcommand{\boron}{\mbox{${}^8$B\ }}

\title{Solar neutrino experiments and Borexino perspectives}
\author{P. Aliani\address[Milano]{Dip. di Fisica, 
        Universit\`a di Milano e I.N.F.N., Sezione di Milano, 
        Via Celoria 16, Milano, Italy},%
        {\underline{V. Antonelli}}\addressmark\thanks{talk given by V. 
         Antonelli},%
          M. Picariello\addressmark,%
         E. Torrente-Lujan\addressmark[Milano]
        \address[MADRID]{Dept. Fisica Teorica,C-XI, 
        Univ. Aut. de Madrid, Spain and CERN TH-Division, 
        CH-1202 Geneve}}
\begin{document}
\begin{abstract}
\vspace{-0.2 cm}

We present an updated analysis of all the data available about solar neutrinos,
including the charged current SNO results. The best fit of the data is 
obtained in the Large Mixing Angle region, but different solutions are still
possible. We also study the perspectives of Borexino and conclude that this 
experiment, with a parallel analysis of total rate and day-night asymmetry, 
should be able to discriminate between the different possible solutions.
\end{abstract}
\maketitle
\vspace{-0.7 cm}
\section{INTRODUCTION}
Fifty years after neutrino discovery, many of its properties (like 
its mass) are not completly understood \cite{torrente1}. 
Many experiments tried to answer these questions using different techniques.
Here we focus our attention on solar neutrinos. 
The SNO experiment \cite{SNO} measured the {\boron} Solar neutrinos through 
the two reactions: (1) charged current (CC): $\nu_e + d\rightarrow 2 p+e^-$ and (2) elastic scattering (ES):  $\nu_x + e^-\rightarrow \nu_x+e^-$.   
The results \cite{SNO} confirmed the previous evidences \cite{solarivecchi} that the flux of $\nu_e$ reaching the Earth is less than the Solar Standard Model (SSM) prediction \cite{SSM,BPB2001}. 
Comparing the two channels, SNO also gave a strong confirmation 
of the validity of $\nu$ oscillation hypothesis and it can be 
considered the first demonstration of the appearance of muon and tau
neutrinos detected at the Earth.    
 In this work we restrict  the analysis to the bidimensional case for 
simplicity,but the extensions to more than two flavours will be treated 
elsewhere \cite{futuro}.
The interest in  neutrino physics is justfied not only by the data  
available, but also by the well founded hope that the forthcoming experiments (like Borexino \cite{Borex} and the long baseline experiments) will be able to 
discriminate more clearly between the possible solutions of the solar neutrino
 problem \cite{strumiavissani}.  
The first aim of our work is to produce a phenomenological analysis 
of all the available solar neutrino data.
We determine the values of 
the mixing parameters compatible with the data and compare the allowed 
regions with the ones selected from Borexino, depending on the 
signal it will find.
The analysis can be divided in the following steps.
We first compute exactly, using an evolution operator formalism
\cite{torrente2}, the {\em survival probability} that a 
neutrino produced with a well determined flavour is still of the same kind 
or has changed flavour when it arrives at the detector.
To take into account the interaction with the 
Earth, we assume a spherical model \cite{modelloterra} in which the Earth is 
divided in eleven radial density zones. 
The other building block of the analysis is the study
of the different aspects of each experiment, as the cross section for the 
interaction of the neutrino in the detector \cite{cross sections}, 
the detector resolution and its efficiency.
More information about this and other points of analysis are reported in 
\cite{noilungo}.
We obtain a {\em response function} for every experiment.
From the convolution of survival probability, {\em response function} 
and $\nu$ flux we obtain the expected signal for every experiment and the 
ratio 
between this value and the one predicted by the SSM in absence of oscillations.
\vspace{-0.425 cm}
\section{STATISTICAL ANALYSIS AND RESULTS}
\vspace{-0.25 cm}
For an exhaustive description of the statistical analysis and of our results 
we refer the interested reader to \cite{noilungo}. Here we just report the 
salient points. 
In the most simple case, one includes in the $\chi^2$ analysis only 
the values of the global rates for all the experiments. The global 
$\chi^2$ function is simply defined as: 
\begin{eqnarray}
\vspace{-0.6 cm}
  \chi^2_{\rm gl}&=& 
({\mbox{\boldmath $R$}}^{\rm th}-{\mbox{\boldmath $R$}}^{\rm exp})^T 
\left(\sigma^{2}\right )^{-1}({\mbox{\boldmath $R$}}^{\rm th}-
{\mbox{\boldmath $R$}}^{\rm exp})
\label{chi1}
\end{eqnarray}
\vspace{-0.5 cm}
where the covariance matrix $\sigma$ is made up by a diagonal part 
(theoretical, statistical and uncorrelated errors) and another part 
(correlated systematic uncertainties).
The ${\mbox{\boldmath $R$}}^{\rm th,exp}$ vectors contain
the data normalized to the SSM expectations:
$R^{th,exp}_i=S_i^{th,exp}/S_i^{SSM},$
where the index $i$ denotes the different Solar experiments:
 Chlorine (Cl), Gallium (Ga), SuperKamiokande (SK) and
 charged current SNO (CC-SNO).
The correlation matrices, both including and excluding SNO, 
 are computed using standard techniques \cite{correlazioni}.
We perform a minimization of the $\chi^2_{gl}$ as a function of the 
oscillation parameters. 
A point in parameter space $(\Delta m^2,\tan^2\theta)$ 
 is allowed if the globally subtracted $\chi_{gl}^2$ fulfills the condition: 
 $\chi_{gl}^2 (\Delta m^2, \theta)-\chi_{min}^2<\chi^2_n(CL)$.
Where $\chi^2_{n=2}$  are the $n=2$ degrees of freedom quantiles.

\begin{center}
\vskip -0.8cm                           
\begin{figure}[htb]
\hskip -0.5cm                             
\begin{tabular*}{2cm}{ll}
\includegraphics*[scale=0.425]{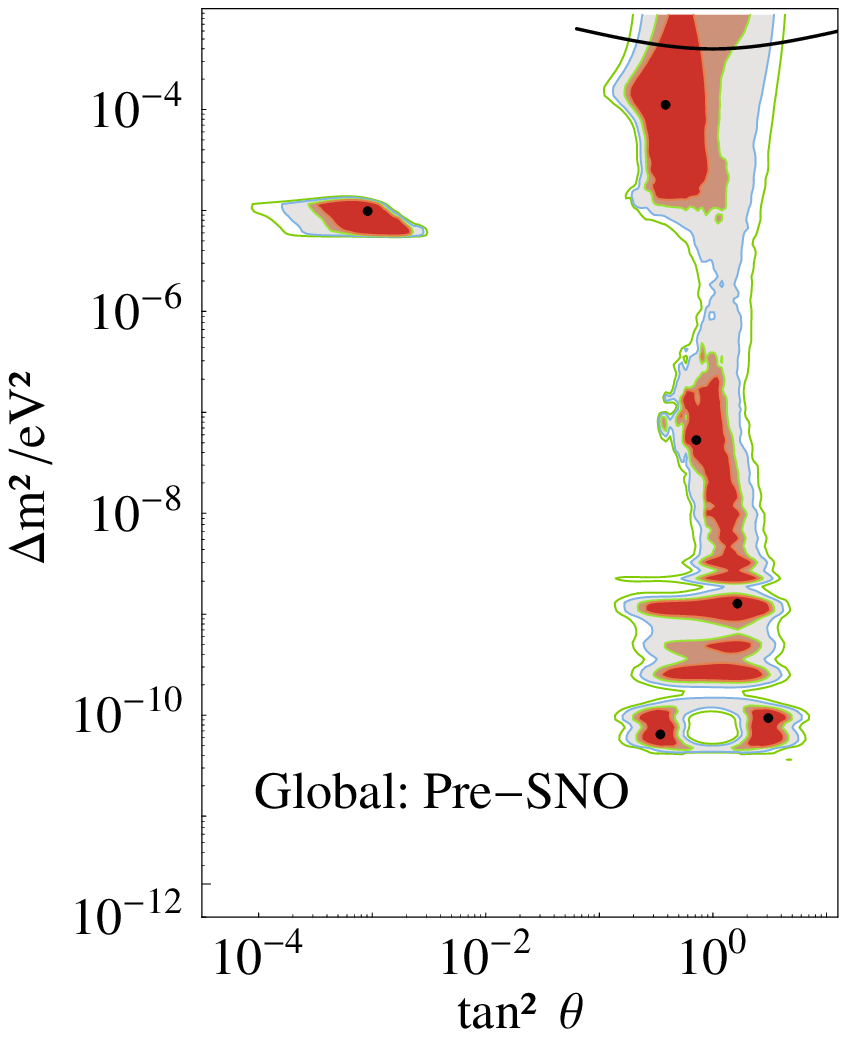}
&\hskip -0.5cm                          
\includegraphics*[scale=0.425]{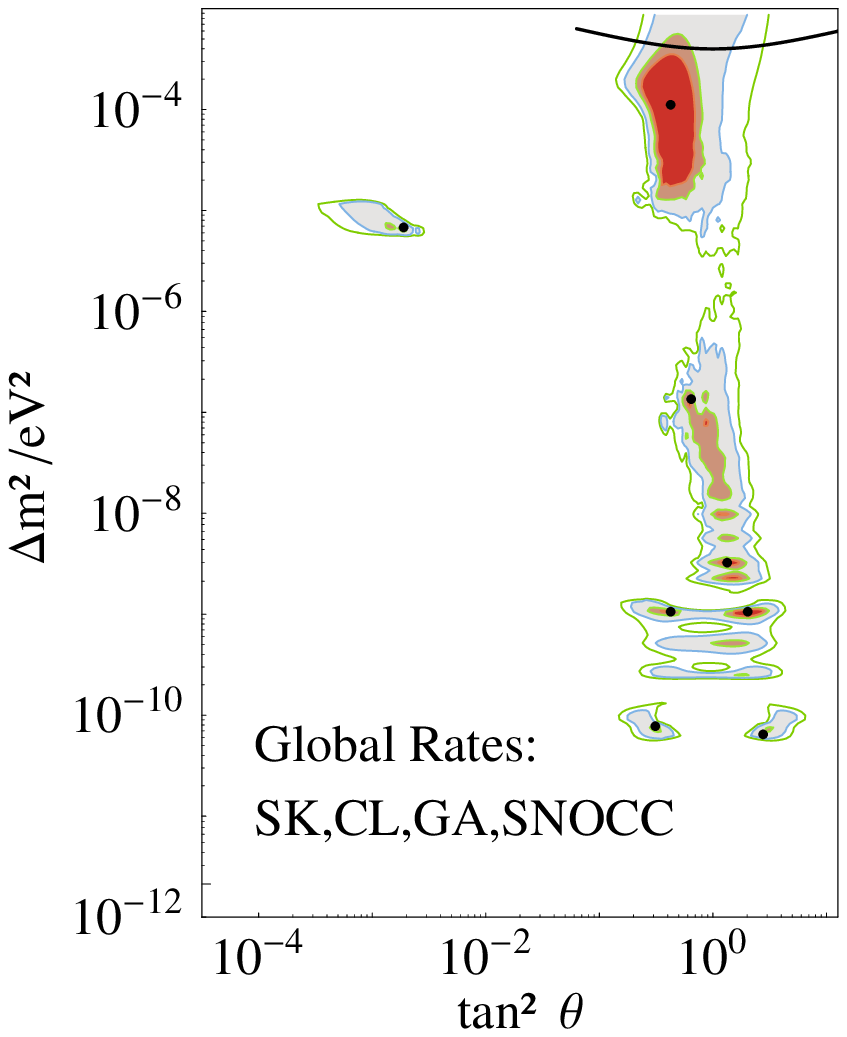}
\end{tabular*}
\vskip -0.8cm                           
%
\vspace{-0.48 cm}    |
\caption{\small (Left) Global rate analysis from SK,Cl and Ga experiments. 
(Right) Same analysis including CC-SNO. The black dots are the best fit 
points; the coloured areas are the allowed regions at 90,95,99 and 99.7 $\%$ 
CL. The region above the solid line is excluded by CHOOZ results at 99 $\%$ 
CL \cite{Chooznew}.}
 \label{global}
\end{figure}
\vskip -1.3cm                          
\end{center}

\vspace{-0.12 cm}
The results of this analysis are represented in Fig. \ref{global}. 
One can distinguish 4 different regions: Small Mixing 
Angle (SMA), Large Mixing Angle (LMA), Low mass (LOW) and 
Vacuum region (VAC). After the introduction of SNO data the different regions 
become well separated.
The best-fit point is no longer the SMA solution, but the one 
in the LMA region and the statistical significance of the SMA region is 
drastically reduced.   
When introducing also the data of the SK energy spectrum rates the statistical
 analysis becomes more complex. In this case we have 41 experimental data 
inputs: the 2x19 values of the bins in which the day and night spectrums are 
divided, plus the total rates for Cl, Ga and CC-SNO. 
The procedure we adopt to define the $\chi^2$ parameter and perform the 
minimization (see \cite{noilungo}) follows the one used by 
the SK collaboration. 
In Table \ref{tabtot} we report the results that one gets in the case in which 
the {\boron} flux is constrained to vary around the BPB2001 \cite{BPB2001} 
central value with the standard deviation given by SSM. The corresponding 
contour plots are 
drawn in Fig. \ref{Borexino} together with the Borexino contour lines. 
The other possibility (free {\boron} flux) is discussed in \cite{noilungo}.   
\begin{table*}[htb]
\vspace{-0.45 cm}
\caption{\small{Best fit oscillation parameters. The analysis includes the 
global rates for Cl,Ga and CC-SNO, and the SK day and night energy spectra.
The flux normalization is constrained to vary with SSM  standard error and 
the number of d.o.f. is $n=41-4$. Also reported are the values of $\chi^2$ 
minimum per degree of freedom ($\chi_{min}^2/n$) and the statistical 
significance (goodness of fit g.o.f.)}}
\label{tabtot}
\newcommand{\cc}[1]{\multicolumn{1}{c}{#1}}
\renewcommand{\tabcolsep}{2pc} 
\renewcommand{\arraystretch}{1.2} 
\begin{center}
\begin{tabular}{@{}lllll}
\hline
Sol & $ \Delta m^2 $ & $\tan^2(\theta)$ &  
$\chi_{min}^2/n$ & g.o.f \\
\hline
LMA        &  5.2 $\times 10^{-05}$ &  0.47  &   0.8 &  77 \\ 
LOW        &  9.9 $\times 10^{-09}$ &  1.03  &   0.9 &  65 \\ 
LOW        &  3.6 $\times 10^{-08}$ &  0.97  &   0.9 &  60 \\ 
VAC        &  5.0 $\times 10^{-10}$ &  1.86  &   1.1 &  28 \\ 
VAC        &  5.0 $\times 10^{-10}$ &  0.52  &   1.1 &  24 \\ 
SMA        &  5.6 $\times 10^{-06}$ &  1.32 $\times  10^{-3}$ &   1.4 &   3.2 \\\hline
\end{tabular}\\[2pt]
\end{center}
\end{table*}
\vspace{-0.35 cm}
\section{BOREXINO PERSPECTIVES}
\begin{center}
\vskip -0.9cm                           
\begin{figure}[htb]
\hskip -0.5cm                           
\begin{tabular*}{2cm}{ll}
\includegraphics*[scale=0.425]{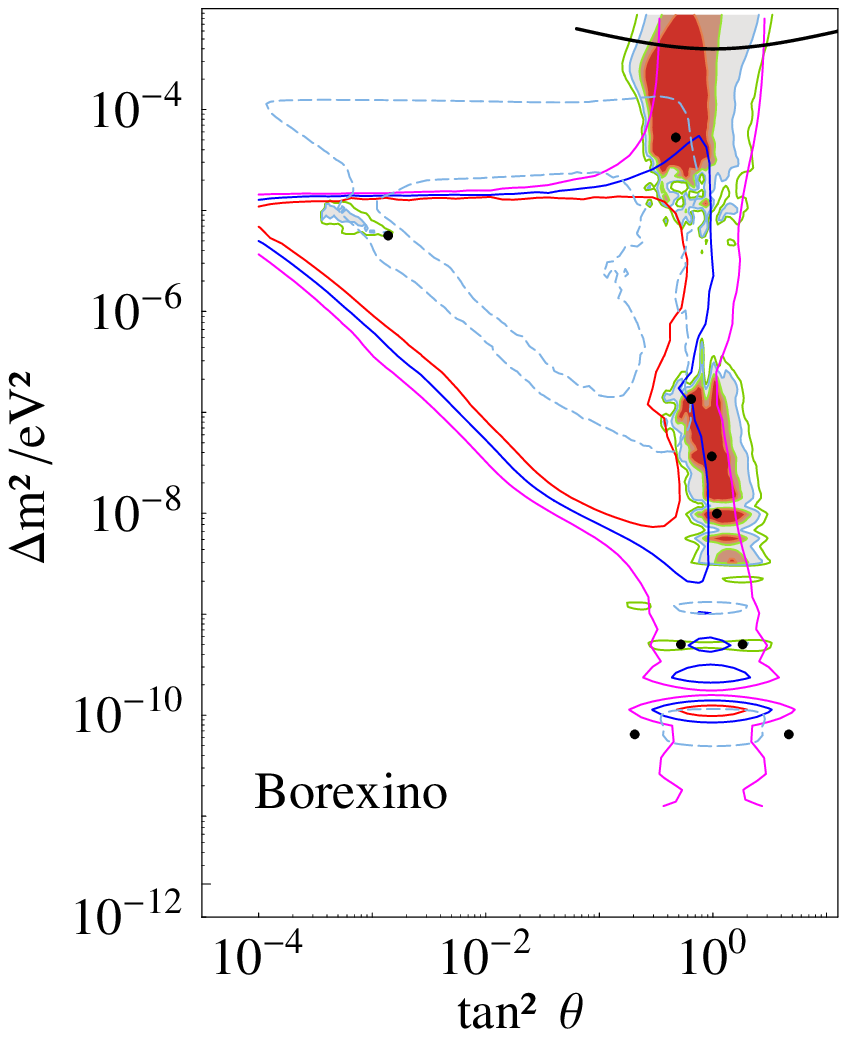}
&\hskip -0.5cm                          
\includegraphics*[scale=0.425]{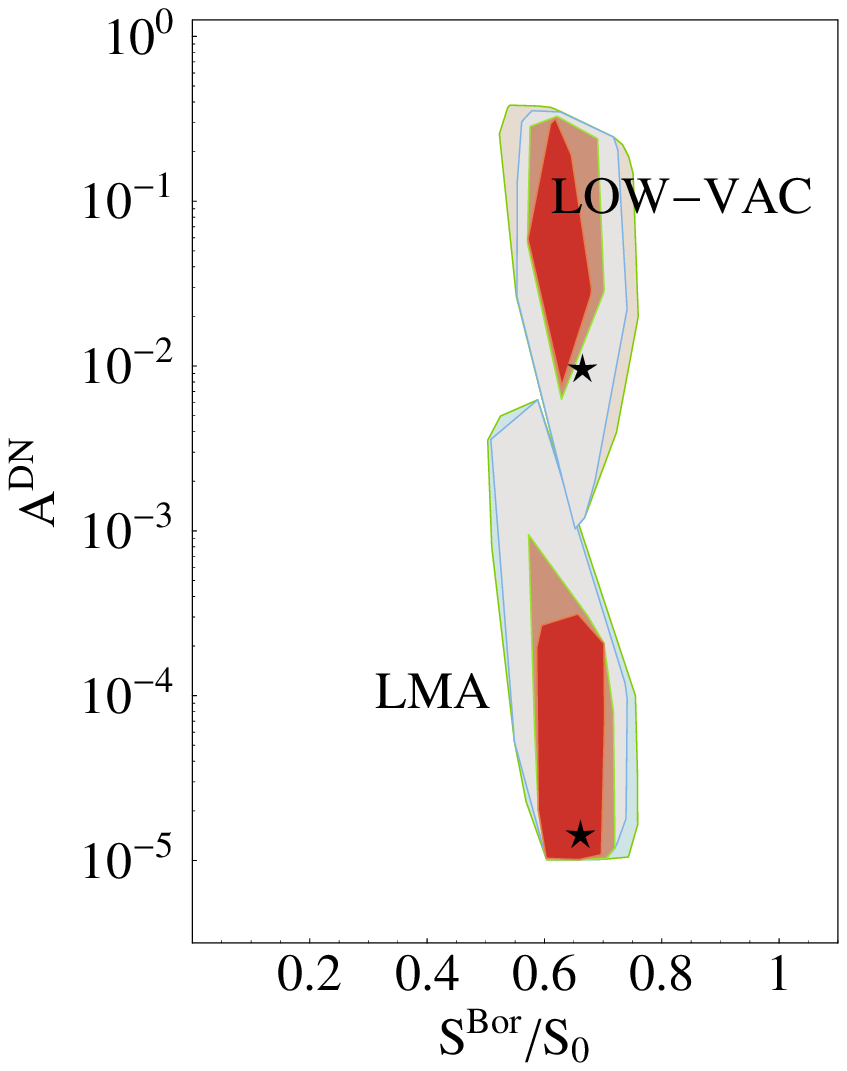}
\end{tabular*}
\vskip -0.9cm                           
\vspace{-0.2 cm}
\caption{\small{(Left) Contour lines (full lines) for 
$S^{Bor}/S_0=0.5, 0.6, 0.7$ superimposed to the contour plots obtained 
from all other experiments and to the regions from Cl experiment alone 
(inside the dashed lines).(Right) Borexino day night asymmetry ($A^{DN}$) 
versus normalized signal ($S^{Bor}/S_0$).}}  
\label{Borexino}
\end{figure}
\vskip -1.3cm                          
\end{center}

\vspace{-0.5 cm}

We analize the results from all other 
experiments together with the expectations for Borexino day-night averaged 
signal normalized with respects to SSM 
($S^{Bor}/S_0=S^{D-N}/S_{SSM}$)          
and day-night asymmetry 
($A^{DN}=2 (D-N)/(D+N)$). 
In Fig. \ref{Borexino} the contour lines corresponding to 
different possible values of 
$S^{Bor}/S_0$ 
are superimposed on the allowed regions obtained from the global analysis of the full set of data for the other experiments.  
The signal discrimination power of the experiment \cite{giamma} should be 
sufficient to distinguish between the different allowed regions in the 
$(\Delta m^2,\tan^2\theta)$ 
plane or at least to strongly favour one of them.
Borexino potentiality becomes even more evident when we look at the day night 
asymmetry. From the second graph of Fig. \ref{Borexino} we see
that the LMA and the LOW regions correspond to quite different values of 
$A^{DN}$. 
\vspace{-1.58 cm}
\section{CONCLUSIONS}
\vspace{-0.35 cm}
We analyzed all the Solar neutrino data available
and  for the best solutions which fit the present data, we studied 
Borexino expectations.
In the most comprehensive case, global rates plus spectrum,
 the best fit was obtained in the LMA region.
Solutions in the LOW and VAC regions are  still possible although
 much less favoured. The best possible solution in the SMA region gets a 
low statistical significance.
From the study of the expected Borexino normalized signal 
and day night asymmetry we conclude the following.
In the near future, after 2-3 years of data taking, 
 the  combined Borexino measurements of the 
 total event rate with an error below $\pm 5-10\%$ and day-night total 
 rate asymmetry with a precision comparable to that of SK should  
 allow us to discriminate between the Solar neutrino solutions suggested by 
present data. 
\vspace{-0.42 cm}
\section*{ACKNOWLEDGMENTS}
\vspace{-0.39 cm}
We are really glad to thank R. Ferrari for his continuous scientific and organizative support, that has been essential for this work. We thank the Milano Borexino group, M. Pallavicini and the 
organizers of TAUP meeting. 
We acknowledge the financial support of MIUR and CYCIT.  

\end{document}